# A kinetic description of Vanadium carbide coating formed by the plasma electrolytic method

**Ramona Javadi Doodran**

M.Sc., Department of Materials Engineering, Amirkabir University of Technology, Tehran, Iran

**Abstract:**

The old creation method of this coating is using salt bath for an approximate time of 6 to 10 hours, which means wasting a lot of time and energy while the formation of this coating with the plasma electrolytic method just takes about 15 minutes and it can increase the efficiency strongly on an industrial scale. In this research, the process has been applied on the samples of 1.2436 steel in different periods. Then, thickness of the coating layers has been measured by using SEM images and thermodynamic data of Vanadium carbide formation. These results have been expanded to access and present a confirmed model to predict the thickness of this diffusion-based coating as a function of time and reaction temperature. The proven model can be used to demonstrate and prove the kinetic advantage of this method and also find the optimal value of applying time and temperature.

**Keywords:** Diffusion; Surfaces; Thermodynamics and kinetics of processes in materials; Metallurgy

1. Introduction

The main goal of surface engineering is looking for new methods to achieve hard coatings as cheap and fast as possible. The plasma electrolysis is one of those novelties which is growing rapidly because of its properties such as low wasting of energy and high rate of process. Our team has already published some scientific papers on coatings via this process cited as references [1-3], those have emphasized on properties of achieved coatings and basically investigation of diffusion based coatings formed by plasma electrolysis besides its merits and demerits for industrial use. In this research, kinetic properties of vanadium carbide coating formed via plasma electrolysis method have been investigated with the diffusion kinetics point of view. Kinetic is one of the difference maker factors among ways of reaching a specific hard coat since in industrial scale, rate of process hugely affects on efficiency and final economic income.[4-6] Therefore a proven kinetic model can help engineers anticipating the optimal time and temperature to achieve vanadium carbide coating as well as demonstrate how much affordable this method is. Essential kinetic data in each step has been extracted from SEM images using MIP4, software which uses the color difference in SEM images to calculate the thickness of coating, and thermodynamic data have been gotten from HSC 5.11 software.

2. Experimental method

The samples used are DIN 1.2436 cold work tool steel with the following chemical compositions (wt.%.): Cr (11-11.5), Mo (0.75-0.8), C (2-2.1), Si (0.2-0.25), and Fe balance. Samples are in cylindrical shape with 20 mm diameter and 22 mm height. To measure the temperature of the sample surface, a hole with 8±0.01mm diameter and 21.87±0.01 mm height was created and a thermocouple was placed inside. Another hole with 3.5±0.01mm diameter and 10±0.01 height to connect the electrode to the sample (Fig.1 A)
The electrolyte consisted of 4±0.1g vanadium oxide, 50±1ml hydro choleric acid, different amounts of sodium hydroxide (NaOH) and water, which was added to reach 2 liters of the solution [3]. All the coating processes were performed in $1253^oK(980^oC)$. Coating voltage was 170±3 v via DC source.
The microscopic observation was carried out by a Philips XL 30 Scanning Electron Microscope and the films composition was investigated by Philips 1840 X-ray diffraction (XRD). The cross-sectioned sample was etched with picral solution ($(NO_2)_3C_6H_2OH$) 5% picric acid and 95%ethanol) for microscopic analysis.

3. Results and Analysis:

To start kinetic analysis, we should initially prove that the coating layer is Vanadium Carbide. To reach this goal, EDX analysis has been applied on the samples. The maps of Vanadium as results of this test have been displayed in Fig.1 C to G. As it is obvious in comparison of figures with SEM analysis images (Fig.1 C to G), the map of Vanadium is approximately coincident with the coating layer area. This is the most accurate and common proof of element diffusion in this low weight percent. Since the results have had good conformity with the diffusion supposition, it can be counted as valid hypothesis and the study can be continued.

The most common available proof of element diffusion is the result of XRD analysis. However, as coating layers were too thin, this test was unable to answer for all of the samples. The low weight percentage of the eligible element is the reason indeed. However, the result of this test on the last sample, which has the most thickness, has been shown in Fig.1 B. It shows that vanadium has diffused and VC has been formed in the coating layer. So many noises in this diagram have been made by irregular reflection of light due to the high surface roughness.

To study the adding rate of thickness at different time periods, SEM analysis was also used. In Fig.1 C to G, the coating layers are completely obvious.

For kinetic analysis, the average thickness of every layer which has been formed is necessary. Consequently, the "Micro structural image processing" software named "MIP4" has been used which measures the thickness of the coating layer, considering the color difference between the surface of base metal and that of the coating layer in the SEM analysis images; and calculates average thickness and standard deviation of all sample images. In Fig.2 A to E, this process has been displayed and the output of this measurement is as Table 1.

As all of the kinetic analysis will be based on the Table 1 information and the STDs are relatively high, to ensure accuracy of the data, hardness profile of the samples for depth of 8,16,24,32,40,48,56, and 80 microns, has been determined. The micro hardness test of Vickers has been applied to the samples. The obtained data is available in Table1.

The data of the table1can be displayed in the following diagram (Fig.2 F) :

According to this diagram, after 6 minutes the sample has been seen to have a drop in hardness at a depth of 8 and 16 microns and it confirms the average amount (12.41 microns) which has been reported before. The sample of 9 minute almost has a constant and high hardness from depth of 24 to 32 microns, in which the reported quantity of 25.17 is also confirmed. The 12 minute sample has its highest level of hardness in depth of 24 microns that is probably due to concentration of Carbide and its collision with the test indicator. This sample also has an acceptable level of hardness for Carbide to depth of 32 microns. However, after that, there has been a high decline in hardness which means the site of coverage has been finished. Thus, the average quantity reported is acceptable in this case too. In the 15 and 18 minute samples, there is not many differences in the hardness profile. Considering the fact that the process is controlled by diffusion, it shows the maximum thickness of the coating layer which is possible to reach. Process completion time and maximum thickness of the coating layer will be fully discussed in the next part. However, both diagrams show a depth of 56 to 80 microns and confirm the reported numbers (18,58,29,59).

According to obtained data reported, the average thickness can be supposed as valid and can be used in the kinetic analysis of coating.

It should be noted that according to the all above mentioned data, the kinetics calculations will be determined with less than 10% deviation which is an acceptable approximation in kinetic problems. As the reaction fraction (α) is used in kinetic equations of solid-state, we must consider a state as a final or complete one of the reaction. Considering the information of table 1, it can be concluded that in relatively long periods of time, the maximum depth of the layer will reach to 60 microns and after that, as time passes, thickness of the coating will not increase. The reason specifically is the limitation of Vanadium diffusion. So, the reaction of Vanadium and Carbon can be proceed to a maximum 60 microns depth of diffusion at this temperature. Thus, by dividing the presented data in table 1 by 60, the amount of reaction fraction can be calculated with a good approximation:

t(s)=6    α=0.206833

t(s)=9    α=0.4195

t(s)=12   α=0.496

t(s)=15   α=0.969667

t(s)=17   α=0.988167

The Excel Software has expanded the data to possible kinetic models. These models contain the limiting ash layer (Ash), limiting chemical reaction (chem.), first order linear, second order linear, third order linear, Avrami, one dimensional diffusion, power rate, first order transfer and second order transfer, etc. The best obtained result is related to the power rate model with a regression radius of $R^2 = 0.957$ (Fig.2 G). Thus considering this number, the model can be chosen as the kinetic model of the process.

According to this diagram (Fig.2 G), the kinetic equation of the reaction is                       . According to the Arrhenius model, the kinetic constant is obtained as below:

$$k = A.exp(\frac{-Q}{RT})$$

We also know that the power rate equation is assumed as the following $\alpha^{\bar{n}} = kt$ . Thus, the Arrhenius equation has been powered by n and has formed the contents of the above equation. So $k_{Arrhenius}=0.058$.

The activation energy of Vanadium Carbide coating is equal to 173.2 KJ/mol [13]. By replacing this number and the amount of R=8.314KJ/mol.°K as well as the reaction temperature of T=1253°K(980°C), the following equation is obtained:

$$0.058 = A \exp(-16.63)$$

$$A = 966041.5$$

The constant amount of A depends on the nature of the reaction and it does not change by the reaction or diffusion conditions. It only depends on term of temperature. Therefore, the general kinetic equation of the Vanadium Carbide coating formation via plasma electrolytic method based on temperature and time of the reaction is as follows:

$$\alpha = [966041.5 \exp(\frac{-20833}{T}) \, t]^{1.4694}$$

$$\alpha = 6.228 \times 10^8 \exp(\frac{-30612}{T}) \, t^{1.4694}$$

In this equation, t is time in minute and T is temperature in Kelvin.

As another result of this equation, it is now possible to compare kinetics of two mentioned methods of Vanadium carbide coating formation :

**Plasma Electrolytic:** $\quad d(\eta m) = 3.73 \times 10^{10} \, [t. \exp(\frac{-20833}{T})]^{147}$

**Salt Bath [6]:** $\quad d(\eta m) = 2.82 \times 10^{-2} \sqrt{t \exp(\frac{-20833}{T})}$

As expected, this comparison demonstrates that the fascinating difference between the two constant numbers causes a great decrease in the time of the process factor in industrial scale by using plasma electrolytic method. It can be used as the basic information to further studies and proposals to suggest replacing traditional methods of hard coating by plasma electrolytic.

4. Conclusion

The obtained data have been expanded to the most famous kinetic models for metal diffusion and eventually, have had the best conformity with the power rate model with a regression radius of 957.

Finally, the formula of the coating layer thickness as a function of time and temperature has been obtained. In comparison with the formula of this coating thickness via the old way, salt bath method, it obviously shows the fact that this coating can be achieved at least $10^{12}$ times faster via the plasma electrolytic.

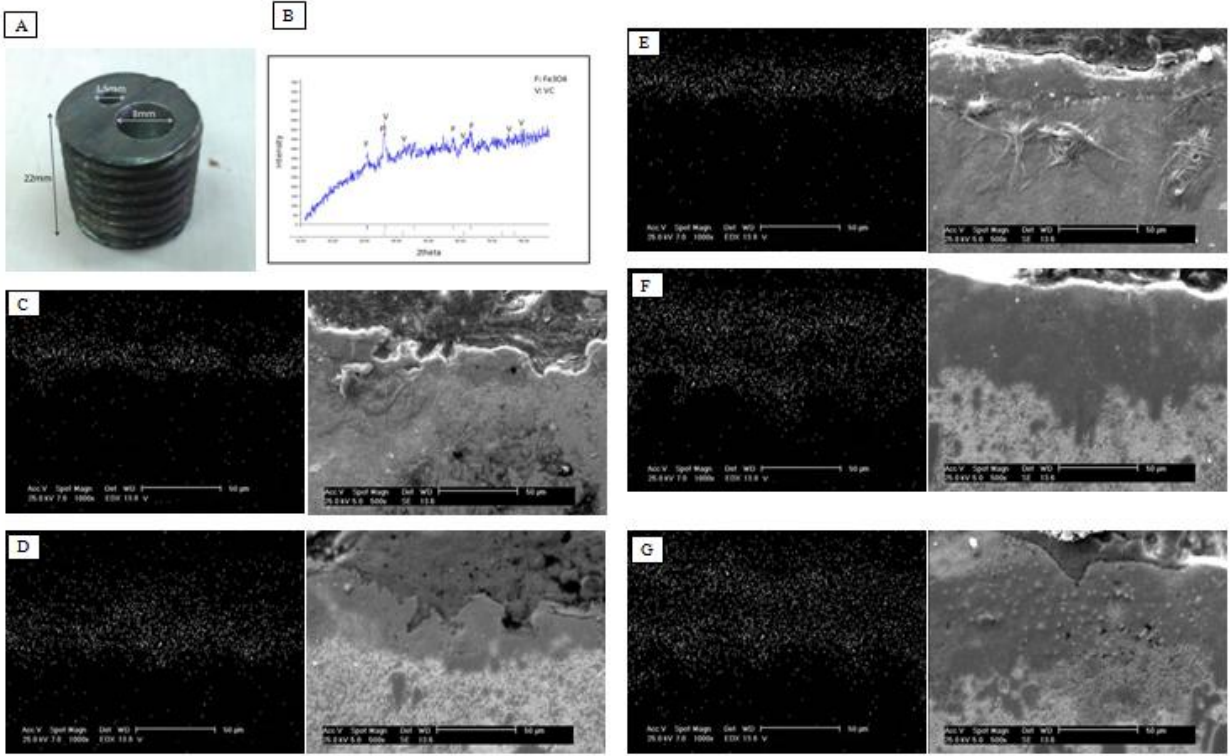

Fig.1

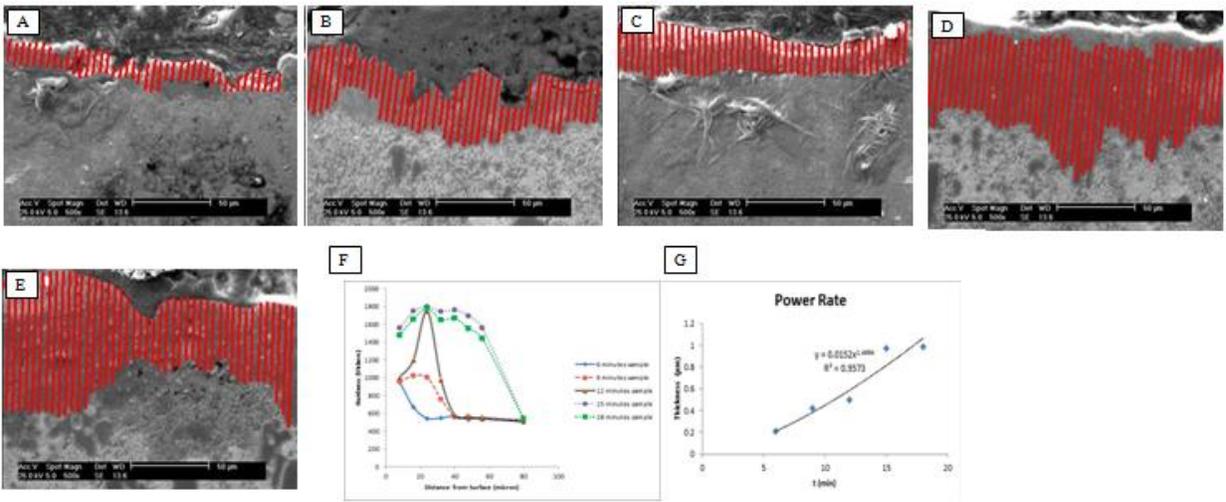

Fig.2

Table.1

| | 8microns | 16microns | 24microns | 32microns | 40microns | 48microns | 56microns | 80microns | Thickness(μm) |
|---|---|---|---|---|---|---|---|---|---|
| 6-minutes-sample | 956±121 | 670±98 | 545±55 | 550±54 | 570±69 | 532±68 | 554±52 | 511±55 | 12.41 |
| 9-minutes-sample | 958±128 | 1020±139 | 1003±166 | 760±101 | 560±57 | 559±59 | 542±59 | 530±61 | 25.17 |
| 12-minutes-sample | 1006±134 | 1200±146 | 1750±170 | 980±112 | 578±64 | 559±61 | 541±57 | 510±59 | 29.76 |
| 15-minutes-sample | 1560±149 | 1750±141 | 1800±164 | 1745±144 | 1763±169 | 1692±166 | 1558±148 | 545±57 | 58.18 |
| 18-minutes-sample | 1480±150 | 1659±139 | 1779±177 | 1653±167 | 1668±165 | 1554±161 | 1442±152 | 535±60 | 59.29 |